%% file: 0main-CSUR.tex
\documentclass[manuscript]{acmart}

\usepackage{algorithm}
\usepackage{algpseudocode}
\usepackage{tikz}
\usepackage{comment}
\usepackage{times}
\usepackage{soul}
\usepackage{url}
\usepackage{enumitem}
\usepackage{wrapfig}
\usepackage{marvosym}
\usepackage{ifsym}
\usepackage{xcolor}
\usepackage{xspace}
\usepackage{booktabs}
\usepackage{multirow}
\usepackage{makecell}
\usepackage[edges]{forest}
\definecolor{hidden-draw}{RGB}{20,68,106}
\definecolor{hidden-pink}{RGB}{255,245,247}

\tikzstyle{my-box}=[
    rectangle,
    draw=hidden-draw,
    rounded corners,
    align=left,
    text opacity=1,
    minimum height=1.5em,
    minimum width=5em,
    inner sep=2pt,
    fill opacity=.8,
    line width=0.8pt,
]

\tikzstyle{leaf-head}=[my-box, minimum height=1.5em,
    draw=gray!80, 
    fill=gray!15,  
    text=black, font=\normalsize,
    inner xsep=2pt,
    inner ysep=4pt,
    line width=0.8pt,
]

\tikzstyle{leaf-task}=[my-box, minimum height=1.5em,
    draw=red!70, 
    fill=red!15,  
    text=black, font=\normalsize,
    inner xsep=2pt,
    inner ysep=4pt,
    line width=0.8pt,
]

\tikzstyle{leaf-paradigms}=[my-box, minimum height=1.5em,
    draw=cyan!70, 
    fill=cyan!15,  
    text=black, font=\normalsize,
    inner xsep=2pt,
    inner ysep=4pt,
    line width=0.8pt,
]
\tikzstyle{leaf-others}=[my-box, minimum height=1.5em,
    draw=orange!80, 
    fill=orange!15,  
    text=black, font=\normalsize,
    inner xsep=2pt,
    inner ysep=4pt,
    line width=0.8pt,
]

\tikzstyle{modelnode-task}=[my-box, minimum height=1.5em,
    draw=red!80, 
    fill=hidden-pink!30,  
    text=black, font=\normalsize,
    inner xsep=2pt,
    inner ysep=4pt,
    line width=0.8pt,
]

\tikzstyle{modelnode-paradigms}=[my-box, minimum height=1.5em,
    draw=cyan!100, 
    fill=hidden-pink!30,  
    text=black, font=\normalsize,
    inner xsep=2pt,
    inner ysep=4pt,
    line width=0.8pt,
]
\tikzstyle{modelnode-others}=[my-box, minimum height=1.5em,
    draw=orange!100, 
    fill=hidden-pink!30,  
    text=black, font=\normalsize,
    inner xsep=2pt,
    inner ysep=4pt,
    line width=0.8pt,
]

\title{Multi-Task Deep Recommender Systems: A Survey}

\author{Yuhao Wang}
\affiliation{%
  \institution{City University of Hong Kong}
  \city{Hong Kong}
  \country{China}
}
\email{yhwang25-c@my.cityu.edu.hk}

\author{Ha Tsz Lam$^{\star}$}
\thanks{$\star$ \text{Equal contribution}}
\affiliation{%
  \institution{City University of Hong Kong}
  \city{Hong Kong}
  \country{China}
}
\email{hatszlam2-c@my.cityu.edu.hk}

\author{Yi Wong$^{\star}$}
\affiliation{%
  \institution{City University of Hong Kong}
  \city{Hong Kong}
  \country{China}
}
\email{ywong692-c@my.cityu.edu.hk}

\author{Ziru Liu}
\affiliation{%
  \institution{City University of Hong Kong}
  \city{Hong Kong}
  \country{China}
}
\email{ziruliu2-c@my.cityu.edu.hk}

\author{Wanyu Wang}
\affiliation{%
  \institution{City University of Hong Kong}
  \city{Hong Kong}
  \country{China}
}
\email{wanyuwang4-c@my.cityu.edu.hk}

\author{Maolin Wang}
\affiliation{%
  \institution{City University of Hong Kong}
  \city{Hong Kong}
  \country{China}
}
\email{Morin.wang@my.cityu.edu.hk}

\author{Yichao Wang}
\affiliation{%
  \institution{Huawei Noah’s Ark Lab}
  \city{Shenzhen}
  \country{China}
}
\email{wangyichao5@huawei.com}

\author{Bo Chen}
\affiliation{%
  \institution{Huawei Noah’s Ark Lab}
  \city{Shanghai}
  \country{China}
}
\email{chenbo116@huawei.com}

\author{Huifeng Guo} 
\affiliation{%
  \institution{Huawei Noah’s Ark Lab}
  \city{Shenzhen}
  \country{China}
}
\email{huifeng.guo@huawei.com}

\author{Ruiming Tang$^{\dagger}$}
\affiliation{%
  \institution{Huawei Noah’s Ark Lab}
  \city{Shenzhen}
  \country{China}
}
\email{tangruiming@huawei.com}

\author{Xiangyu Zhao$^{\dagger}$}
\thanks{$\dagger$ \text{Corresponding authors}}
\affiliation{%
  \institution{City University of Hong Kong}
  \city{Hong Kong}
  \country{China}
}
\email{xianzhao@cityu.edu.hk}

\renewcommand{\shortauthors}{Yuhao Wang et al.}
\begin{abstract}
    Multi-task learning (MTL) aims at learning related tasks in a unified model to achieve mutual improvement among tasks considering their shared knowledge. It is an important topic in recommendation due to the demand for multi-task prediction considering performance and efficiency. Although MTL has been well studied and developed, there is still a lack of systematic review in the recommendation community. To fill the gap, we provide a comprehensive review of existing multi-task deep recommender systems (MTDRS) in this survey. To be specific, the problem definition of MTDRS is first given, and it is compared with other related areas. Next, the development of MTDRS is depicted and the taxonomy is introduced from the task relation and methodology aspects. Specifically, the task relation is categorized into parallel, cascaded, and auxiliary with main, while the methodology is grouped into parameter sharing, optimization, and training mechanism. The survey concludes by summarizing the application and public datasets of MTDRS and highlighting the challenges and future directions of the field.
\end{abstract}

\begin{document}


\maketitle


\input{1Introduction}

\input{2Taxonomy}

\input{3Application}

\input{5FutureDirections}

\input{6Conslusion}

\bibliographystyle{ACM-Reference-Format}
\bibliography{ref_new_2}
\end{document}

%% file: 1Introduction.tex
\section{Introduction} \label{intro}



The development of the Internet industry has led to a tremendous increase in the information volume of online services, such as social media and online shopping platforms \cite{aceto2020industry}. The Recommender Systems (RS), which match different types of items with users based on their hidden patterns, has made significant influences on improving the online experience for users in a variety of scenarios, such as product matching in online shopping and movie recommendation~\cite{ma2018entire,tang2020progressive}.
For recommendation problems, the input data typically consists of categorical features, sparse identity information, and user--item interaction signals \cite{ma2018entire}. The common recommendation tasks can be divided into score prediction and generation. Score prediction is usually formulated as a classification, regression, or ranking problem predicting the likelihood or utility of a user action, such as click through rate (CTR) prediction. Generation-oriented tasks produce recommendation-related text or sequences, such as explanations for recommendations \cite{lu2018like}.


In practice, the RS should be endowed with the capability to conduct various recommendation tasks simultaneously, so as to cater to multiple and diverse demands of user. For example, in video recommendation, users exhibit different behaviors towards a single video, such as clicks, likes, and retweets. Consequently, the development of multi-task recommendation (MTR) is prompted in both research and application fields. To be specific, MTR is applied in various domains for more personalized and relevant recommendations based on multiple aspects of user behaviors. {This survey focuses on deep neural multi-task recommender systems, where multiple recommendation tasks or feedback labels are modeled in a shared deep framework. Classical non-deep MTL recommendation methods are discussed only as background when they help clarify definitions or historical development.} Based on the capability of deep neural networks (DNN) to learn high-order feature interactions together with linear ones and model complex user-item interaction patterns, multi-task deep recommender systems (MTDRS) incorporate multi-task learning (MTL) paradigm and have demonstrated superior performance compared to traditional MTR frameworks.

Compared with tackling multiple recommendation tasks separately, MTDRS  offers two main benefits. On the one hand, by exploiting data and knowledge across multiple tasks, 
MTDRS can achieve mutual enhancement among the tasks. On the other hand, it obtains higher efficiency of computation and storage. 
Despite these advantages, MTDRS also faces three challenges. First, MTDRS must effectively and efficiently capture useful information and relevance among tasks. Second, the data sparsity, e.g., in the conversion signal, presents a challenge. Third, the unique sequential dependency, i.e., the sequential pattern of user actions across tasks in recommendations, is another challenge faced by MTDRS.

\begin{figure*}[t]
\setlength\abovecaptionskip{0.1\baselineskip}
\setlength\belowcaptionskip{0.2\baselineskip}
\centering
\includegraphics[width=0.92\textwidth]{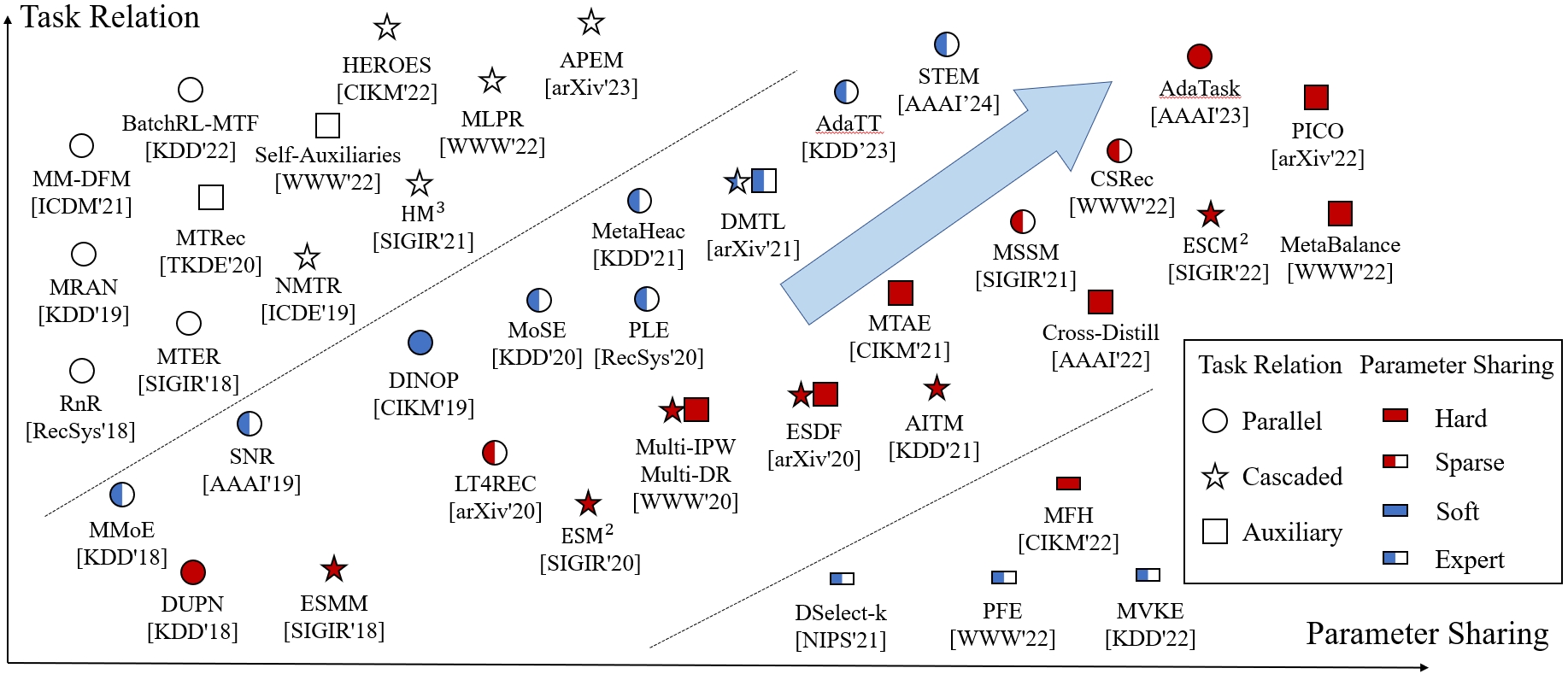}
\caption{{Trend of MTDRS. Shapes denote task-relation categories and colors denote parameter-sharing patterns. The two axes are conceptual coordinates for organizing representative works rather than ordinal performance scales; the large arrow indicates the overall development. Overlaps indicate that one model may belong to both dimensions.}} 
\label{trend}
\vspace{-6mm}
\end{figure*}

In the following, the problem definition and analysis of MTDRS are presented in Section \ref{prel}, including the comparison with three similar areas in recommendation and MTL in CV and NLP. Next, the trend (Figure \ref{trend}) and the taxonomy of MTDRS are detailed in Section \ref{taxo}. Afterward, the application of MTDRS and public datasets are introduced in Section \ref{app}. Finally, the challenges and future directions of MTDRS are summarized in Section \ref{fd} and Section \ref{concl} concludes the survey.

\section{Problem Definition and Analysis} \label{prel}
In this section, the formulation and loss function of MTDRS are given. Next, MTR is compared with related fields in recommendation, and the MTL techniques in CV and NLP areas.

\subsection{\textbf{Formulation of MTDRS}}

Given a $K$-task recommendation dataset with $D:= \left\{\textbf{x}_n, (y_n^1,...,y_n^K)\right\}_{n=1}^N $, which consists of $N$ user-item interaction records and the corresponding feature vector of observed impression, where $\textbf{x}_n$ represents the concatenation of the $n$-th user-item ID pair and feature vector.
Each data record has $K$ labels ${y^1,\cdots,y^K}$ for the corresponding tasks. {For example, in short-video recommendation, $\textbf{x}_n$ may include user ID, video ID, device, time, scenario, user profile, item attributes, and recent behavior sequence. A two-task setting can set $y_n^1=1$ if the user clicks the video and $y_n^2=1$ if the user likes it after exposure; a richer setting can further include watch completion, comment, share, or follow labels. These labels correspond to different but related recommendation tasks on the same impression record.} The objective is to learn the MTL model with task-specific parameters $\{\theta^1,...,\theta^K\}$ and shared parameter $\theta^s$, which outputs the $K$ task-wise predictions by extracting the hidden pattern of user-item feature interactions. The loss function for multi-task training is commonly defined as a weighted sum of task losses. 
The above modeling process can be written as the following optimization problem:
\begin{equation} \label{eq:iwl}
    \begin{aligned}
        \underset{\left\{\theta^1,...,\theta^K \right\}}{\arg \min} \mathcal{L}(\theta^s, \theta^1,\cdots,\theta^K) = \underset{\left\{\theta^1,...,\theta^K \right\}}{\arg \min} \sum_{k=1}^{K} \omega^k L^k(\theta^s, \theta^k)
    \end{aligned}
\end{equation}
where $L^k(\theta^s, \theta^k)$ is the loss function for the $k$-th task with parameter $\theta^s,\theta^k$, and $\omega^k$ is the loss weight for the $k$-th task.

\begin{itemize}[leftmargin=*]
    \item The objective functions for most existing MTL works are typically linear scalarizations of the multiple-task loss functions, which fix the weight with a constant. However, the PLE model \cite{tang2020progressive} proposes an updatable loss weight: given an initial loss weight $\omega^k_0$ for task $k$, the loss weight is updated based on a constant ratio $\gamma_k$:
    \begin{equation}
        \omega^k_t = \omega^k_0 . \gamma_k^t
    \end{equation}
    This setting is based on the observation that tasks may have different importance for the specific training period.
    \item Since MTDRS usually conducts score prediction task, one common choice of loss function $L^k(\theta^s,\theta^k)$ for the $k$-th binary feedback task with parameter $\theta^s,\theta^k$ is the BCE loss:
    \begin{equation}
        L^k(\theta^s,\theta^k)= -\sum_{n=1}^N [y_n^k log(\hat{y}_n^k) + 
        \\ (1-y_n^k) log(1-\hat{y}_n^k)]
    \end{equation}
    where $\hat{y}_n^k$ is the prediction value for task $k$ at the $n$-th data parameterized by $\theta^s,\theta^k$. {BCE is not the only possible objective. Depending on the task and feedback type, $L^k$ can also be a regression loss for dwell time or rating prediction, a pairwise or listwise ranking loss for top-$N$ recommendation, a reconstruction loss for representation learning, or a sequence-generation loss for generative recommendation. Thus, Equation \ref{eq:iwl} should be understood as a general scalarized multi-task objective rather than a restriction to BCE-based implicit-feedback prediction.}
\end{itemize}


\subsection{Comparison with Related Recommendation Directions}
\begin{wrapfigure}{r}{0.6\textwidth}
\vspace{-10pt}
\setlength\belowcaptionskip{0\baselineskip}
\setlength\abovecaptionskip{0\baselineskip}
	\centering
	\includegraphics[width=0.98\linewidth]{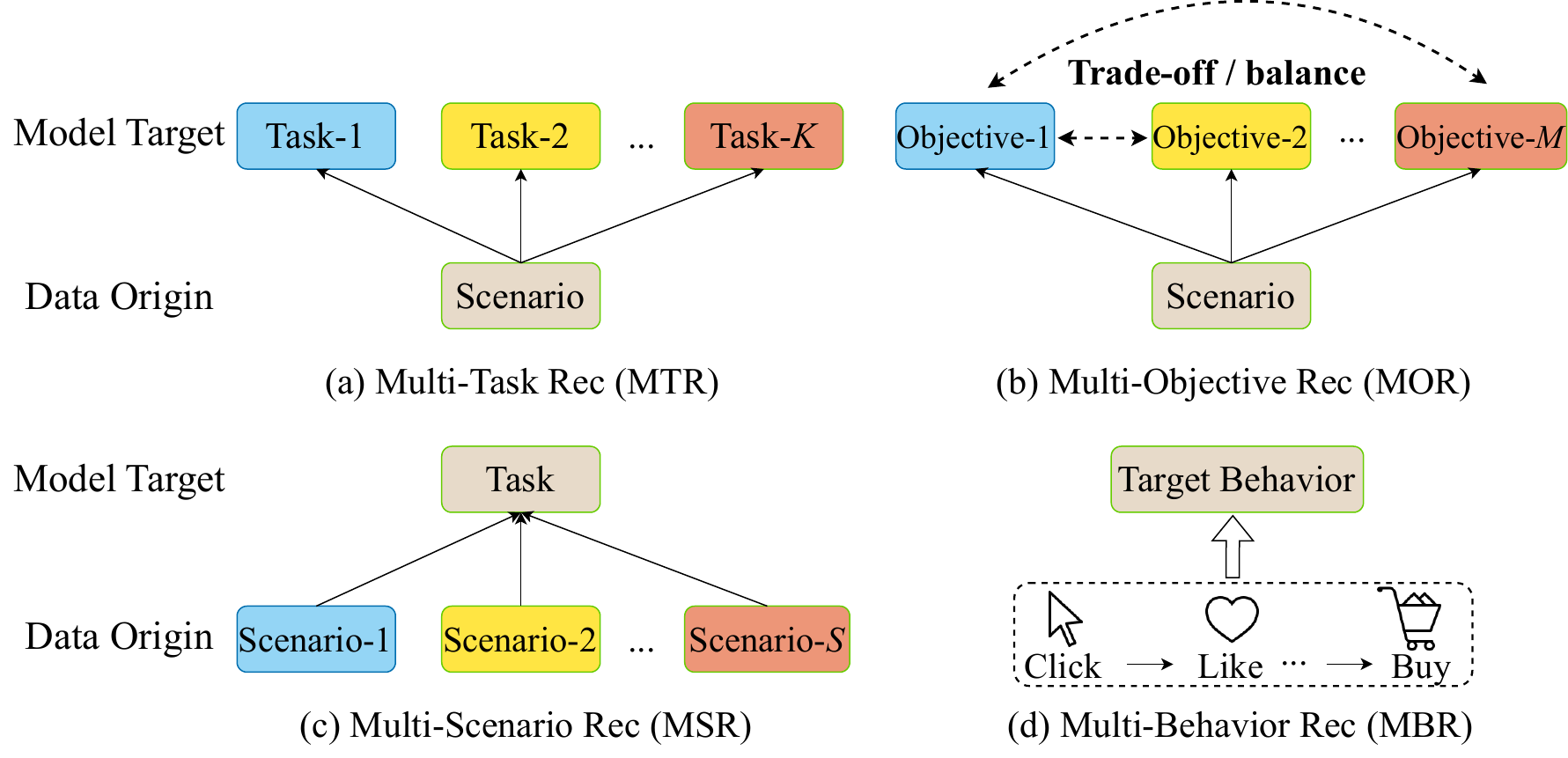}
	\caption{Comparison of MTR, MOR, MSR, and MBR according to the target and data. $K$, $M$, and $S$ denotes the number of task, objective, and scenario.}
	\label{comparison}
    \vspace{-10pt}
\end{wrapfigure}

This section aims to compare the MTR against related areas in recommendation, i.e.,  multi-objective recommendation (MOR), multi-scenario recommendation (MSR) and Multi-behavior recommendation (MBR).
Meanwhile, their comparison is depicted in Figure \ref{comparison}.

\subsubsection{Comparison with Multi-objective Recommendation}

Multi-objective optimization (MOO) studies optimization problems with multiple objectives that cannot generally be reduced to a single preference-free optimum \cite{sawaragi1985theory}. Pareto efficiency is a common optimality criterion, describing a state where no objective could be improved without hurting others. The Pareto frontier is constructed by combing all Pareto-efficient situations.
Based on MOO, multi-objective recommendation (MOR) focuses on the trade-off and balance among multiple objectives, e.g., diversity \cite{liu2021diversity} and fairness \cite{xiao2017fairness} from the optimization perspective. 
Interested readers can refer to the existing surveys on MOR \cite{jannach2022multi,zheng2022survey}.
Furthermore, the connections between MOR and MTR are two-fold. 
On the one hand, MOO can be a solution to MTDRS, since minimizing the losses of different tasks in multi-task learning can be formulated as MOO problem \cite{sener2018multi}, e.g., Lin et al. \cite{lin2019pareto} achieve improvement simultaneously on CTR and GMV task in the online experiment. On the other hand, MTDRS can itself be deployed under multiple business objectives, where each task loss may interact with objectives such as revenue, engagement, fairness, or satisfaction. We further discuss these task--objective interactions in Section \ref{mt}.

\subsubsection{Comparison with Multi-scenario Recommendation}

Multi-scenario recommendation (MSR) aims at improving recommendation performance simultaneously in multiple scenarios such as the different categories of products on Amazon and different spots to present items on Taobao. MSR uses all-scenario data with a unified model to tackle the data sparsity problem. It is also referred to as multi-target cross-domain recommendation in the research community. Similarly, both MSR and MTR require generating adaptive representations among multiple scenarios or tasks in a fine-grained manner and modeling complex inter-dependency. Meanwhile, MTDRS models can be adapted for MSR by treating scenarios as scenario-specific heads or gates over shared representations, as in HMoE \cite{li2020improving}. However, these two problems are at different levels. Specifically, MSR focuses on the same task whose label space is the same while the data distribution is distinctive in different scenarios. By contrast, MTDRS has different label spaces of different tasks. {Therefore, pure MSR studies are outside the core scope of this survey unless they explicitly combine multiple scenarios with multiple recommendation tasks.}

\subsubsection{Comparison with Multi-behavior Recommendation}

Multi-behavior recommendation (MBR) focuses on predicting the probability that a user will interact with an item under the target behavior as single-task learning, given the multi-behavior interaction history. It is inspired by the fact that other types of behavior and feedback contain mutual inter-dependency and contextual signals, leading to a better understanding of target behavior. Similarly, the prediction of different behaviors can be explicitly modeled as different tasks in MTDRS \cite{gao2019neural,ding2021mssm,he2022metabalance}. {The key distinction is that MBR may use multiple behaviors only as input signals for one target behavior, whereas MTDRS optimizes multiple behavior-label prediction or generation tasks jointly.}


\subsection{Comparison with MTL in CV\&NLP}
By contrast, the focus of MTL is different in computer vision (CV) and natural language processing (NLP). {Specifically}, CV MTL often studies dense perception tasks such as detection, segmentation, depth estimation, and classification, where tasks share visual backbones but differ in spatial output structures. Most existing methods utilize feature transformation \cite{zhang2021survey} to represent common features based on a multi-layer feed-forward network. 
Besides, the existing MTL models in NLP study not only architectures but also training objectives, task sampling, transfer, and prompt or sequence-to-sequence formulations \cite{chen2021multi}. {This comparison is useful because MTDRS inherits general MTL issues such as negative transfer and task balancing, but it differs from CV/NLP in the prevalence of sparse categorical features, delayed and biased feedback, industrial ranking constraints, and behavior dependencies such as impression--click--conversion.}

%% file: 2Taxonomy.tex
\section{Taxonomy} \label{taxo}

\begin{figure*}[t]
\setlength\belowcaptionskip{0.1\baselineskip}
\setlength\abovecaptionskip{0.1\baselineskip}
\centering
\includegraphics[width=0.98\textwidth]{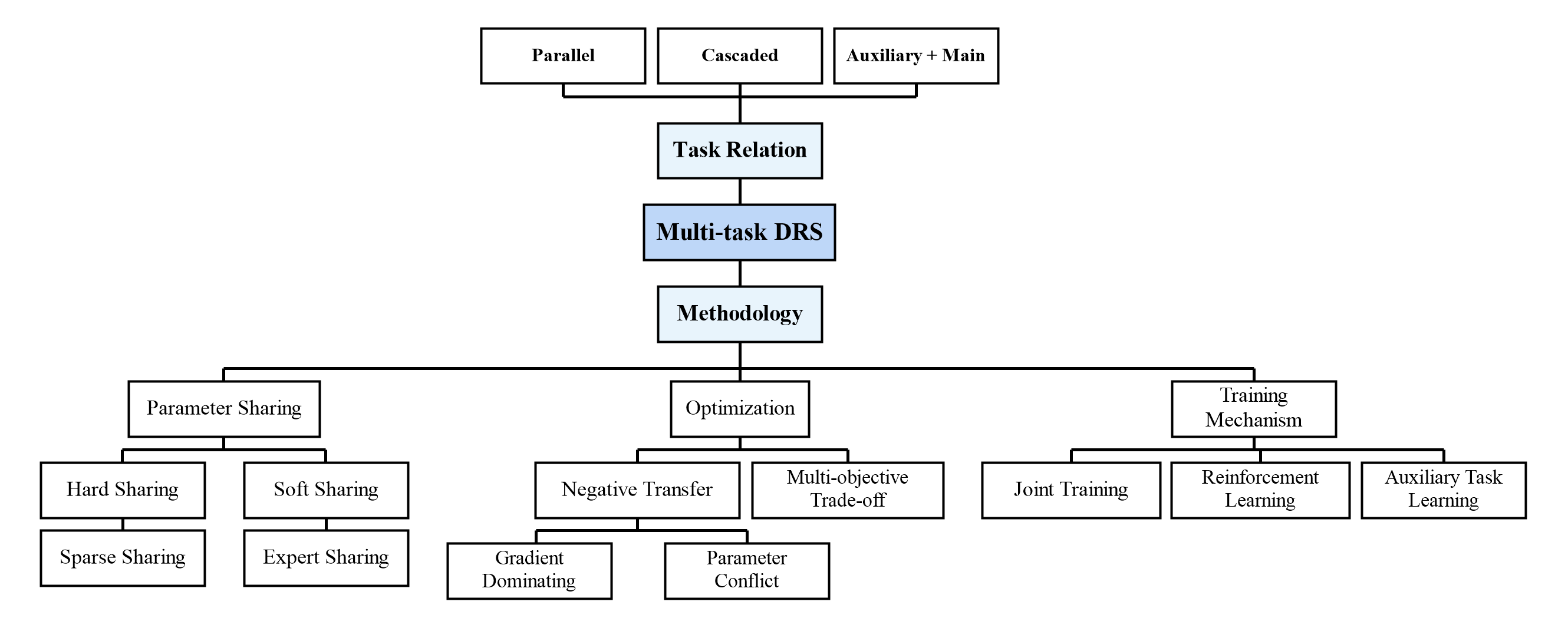}
\caption{{Overview of the facet-based taxonomy. Task relation and methodology are complementary dimensions, so the same MTDRS model can be categorized under both dimensions.}}
\label{tax}
\vspace{-3mm}
\end{figure*}

In this section, the trend of MTDRS is first depicted in Figure \ref{trend}. Next, an overview of taxonomy and detailed taxonomy are depicted in Figure \ref{tax} and \ref{fig_taxonomy}, then it is further illustrated from the perspective of task relation and methodology. {The taxonomy is organized as a facet-based taxonomy rather than a mutually exclusive tree. Specifically, task relation describes how recommendation tasks are related in labels or behavior dependencies, while methodology describes how a model shares parameters, optimizes objectives, or trains tasks. Therefore, one model may appear in both facets, and even in multiple methodological leaves if it combines, for example, expert sharing and gradient balancing.}

\subsection{Trend of MTDRS}
We differentiate MTDRS models by two dimensions, i.e., task relation and methodology. Specifically, we depict the development of MTDRS in Figure \ref{trend} and existing models are divided into three groups: purely on task relation, purely on parameter sharing, and both. 

{To begin with, as for the methods considering task relation only, parallel and cascaded relations are frequently studied in the papers collected by our survey methodology. Meanwhile, among the methods concerned with parameter sharing only, many works target at improving expert sharing. Finally, many representative MTDRS models simultaneously consider these two factors. Among them, a substantial portion of the collected works focus on hard sharing, which indicates the widespread use of shared bottom architecture \cite{caruana1993multitask}. Besides, expert sharing is a hot spot, while it is mainly discussed under the parallel task relation setting.}

\tikzstyle{my-box}=[
    rectangle,
    draw=hidden-draw,
    rounded corners,
    align=left,
    text opacity=1,
    minimum height=1.5em,
    minimum width=5em,
    inner sep=2pt,
    fill opacity=.8,
    line width=0.8pt,
]

\tikzstyle{leaf-head}=[my-box, minimum height=1.5em,
    draw=gray!80, 
    inner xsep=2pt,
    inner ysep=4pt,
    line width=0.8pt,
]

\tikzstyle{leaf-task}=[my-box, minimum height=1.5em,
    draw=red!70, 
    inner xsep=2pt,
    inner ysep=4pt,
    line width=0.8pt,
]

\tikzstyle{leaf-paradigms}=[my-box, minimum height=1.5em,
    draw=cyan!70, 
    inner xsep=2pt,
    inner ysep=4pt,
    line width=0.8pt,
]
\tikzstyle{leaf-others}=[my-box, minimum height=1.5em,
    draw=orange!80, 
    inner xsep=2pt,
    inner ysep=4pt,
    line width=0.8pt,
]

\tikzstyle{modelnode-task}=[my-box, minimum height=1.5em,
    draw=red!80, 
    inner xsep=2pt,
    inner ysep=4pt,
    line width=0.8pt,
]

\tikzstyle{modelnode-paradigms}=[my-box, minimum height=1.5em,
    draw=cyan!100, 
    inner xsep=2pt,
    inner ysep=4pt,
    line width=0.8pt,
]
\tikzstyle{modelnode-others}=[my-box, minimum height=1.5em,
    draw=orange!100, 
    inner xsep=2pt,
    inner ysep=4pt,
    line width=0.8pt,
]

\begin{figure*}[t]
    \centering
    \setlength\belowcaptionskip{0.2\baselineskip}
    \setlength\abovecaptionskip{0.4\baselineskip}
    \resizebox{1\textwidth}{!}{
        \begin{forest}
            forked edges,
            for tree={
                grow=east,
                reversed=true,
                anchor=base west,
                parent anchor=east,
                child anchor=west,
                base=left,
                font=\large,
                rectangle,
                draw=hidden-draw,
                rounded corners,
                align=left,
                minimum width=1em,
                edge+={darkgray, line width=1pt},
                s sep=3pt,
                inner xsep=0pt,
                inner ysep=3pt,
                line width=0.8pt,
                ver/.style={rotate=90, child anchor=north, parent anchor=south, anchor=center},
            }, 
            [
                Multi-Task Deep Recommender Systems,leaf-head, ver
                [
                     Task Relations \\ (\S\ref{tr}), leaf-task,text width=6.5em
                    [
                        Parallel \\ (\S 3.2.1), leaf-task, text width=5em
                        [RnR \cite{hadash2018rank}{, }MTER \cite{wang2018explainable}{, }CAML \cite{chen2019co}{, }DINOP \cite{xin2019multi}{, }DUPN \cite{ni2018perceive}{, } MRAN \cite{zhao2019multiple}{, } \\ RevMan \cite{li2021revman}{, }MSSM \cite{ding2021mssm}{, }CFS-MTL \cite{chen2022cfs}{, }{TIDM \cite{yan2024multi}{, }ORCA \cite{luo2025orca}}, modelnode-task, text width=47.8em]
                    ]
                    [
                        Cascaded \\ (\S 3.2.2), leaf-task, text width=5em
                        [AITM \cite{xi2021modeling}{, }APEM \cite{tao2023task}{, }ESMM \cite{ma2018entire}{, }{AECM \cite{zhang2024aecm}{, }DDPO \cite{su2024ddpo}}, modelnode-task, text width=47.8em]
                    ]
                    [
                        Auxiliary + Main \\ (\S 3.2.3), leaf-task, text width=7.8em
                        [Multi-IPW \& Multi-DR \cite{zhang2020large}{, }ESDF \cite{wang2020delayed}{, }DMTL \cite{zhao2021distillation}{, }MetaBalance \cite{he2022metabalance}{, }MTRec \cite{li2020multi}{, } \\ PICO \cite{lin2022personalized}{, }MTAE \cite{yang2021multi}{, }Cross-Distill \cite{yang2022cross}{, }CSRec \cite {bai2022contrastive}{, }\cite{wang2022can}, modelnode-task,text width=45.0em]
                    ]
                ]
                [
                    Methodology \\ (\S\ref{sec:3-3}), leaf-paradigms,text width=6.5em
                    [
                        Parameter Sharing \\ (\S 3.3.1), leaf-paradigms,text width=8.2em
                        [
                            Hard Sharing, leaf-paradigms, text width=5.8em
                            [
                                Sparse Sharing, leaf-paradigms, text width=6.6em
                                [LT4REC \cite{xiao2020lt4rec}{, }MSSM \cite{ding2021mssm}{, }CSRec \cite{bai2022contrastive}, modelnode-paradigms,text width=28.8em]
                            ]
                        ]
                        [
                            Soft Sharing, leaf-paradigms, text width=5.8em
                            [Expert Sharing, leaf-paradigms, text width=6.6em
                            [MoE \cite{jacobs1991adaptive}{, }MMoE \cite{ma2018modeling}{, }SNR \cite{ma2019snr}{, }PLE \cite{tang2020progressive}{, }DMTL \cite{zhao2021distillation}{, } \\ DSelect-k \cite{hazimeh2021dselect}{, }MetaHeac \cite{zhu2021learning}{, }PFE \cite{xin2022prototype}{, }MVKE \cite{xu2022mixture}{, } \\ FDN \cite{zhou2023feature}{, }AdaTT \cite{li2023adatt}{, }{STEM \cite{su2023stem}{, }MoME \cite{xu2024mome}}{, }MoSE \cite{qin2020multitask}, modelnode-paradigms,text width=28.8em]
                            ]
                        ]
                    ]
                    [
                        Optimization \\ (\S 3.3.2), leaf-paradigms,text width=5.8em
                        [
                            Negative Transfer, leaf-paradigms, text width=7.9em
                            [
                                Gradient \\ Dominating, leaf-paradigms, text width=5.1em
                                [AdaTask \cite{yang2022adatask}{, }MetaBalance \cite{he2022metabalance}{, }{GradCraft \cite{bai2024gradcraft}}, modelnode-paradigms,text width=30.6em]
                            ]
                            [
                                Parameter \\ Conflict, leaf-paradigms, text width=5.1em
                                [PLE \cite{tang2020progressive}{, }CSRec \cite{bai2022contrastive}{, }PEPNet\cite{chang2023pepnet}{, }{TIDM \cite{yan2024multi}{, }GradCraft \cite{bai2024gradcraft}}, modelnode-paradigms,text width=30.6em]
                            ]
                        ]
                        [
                            Multi-objective \\ Trade-off , leaf-paradigms, text width=6.8em
                            [MTA-F \cite{wang2021understanding}{, }Self-Auxiliaries \cite{wang2022can}, modelnode-paradigms,text width=38.5em]
                        ]
                    ]
                    [
                        Training Mechanism \\ (\S 3.3.3), leaf-paradigms, text width=9.2em
                        [
                            Joint Training, leaf-paradigms,text width=6.6em
                            [M2TRec \cite{shalaby2022m2trec}{, }GCM-SR \cite{qiu2021incorporating}{, }MKM-SR \cite{meng2020incorporating}{, }MARRS \cite{das2022marrs}{, }MKR \cite{wang2019multi}{, } \\ \cite{lu2018like}{, }\cite{wang2018explainable}{, }M2GRL \cite{wang2020m2grl}{, }CSRec \cite{bai2022contrastive}{, }{CKF \cite{zhao2025ckf}}, modelnode-paradigms,text width=35.3em]
                        ]
                        [
                            Reinforcement \\ Learning, leaf-paradigms, text width=6.6em
                            [BatchRL-MTF \cite{zhang2022multi}{, }DRRS \cite{han2019optimizing}{, }RMTL \cite{liu2023multi}{, }IntegratedRL-MTF \cite{liu2024off}{, }\\ {EnhancedRL \cite{liu2024enhancedrl}{, }xMTF \cite{cao2025xmtf}}, modelnode-paradigms,text width=35.3em]
                        ]
                        [
                            Auxiliary Task \\ Learning, leaf-paradigms, text width=6.6em
                            [ESDF \cite{wang2020delayed}, modelnode-paradigms,text width=35.3em]
                        ]
                    ]
                    [
                        {LLM / Generative} \\ {Unification}, leaf-paradigms, text width=8.4em
                        [{CKF \cite{zhao2025ckf}{, }URM \cite{jiang2025urm}{, }IDGenRec \cite{tan2024idgenrec}{, }COBRA \cite{yang2025cobra}{, }UniGRF \cite{zhang2025unigrf}}, modelnode-paradigms,text width=44.5em]
                    ]
                ]
            ]
        \end{forest}
    }
    \caption{{Detailed taxonomy of research on Multi-task DRS. The branches summarize representative model placements under each facet and are not intended to be mutually exclusive model partitions.}} 
    \label{fig_taxonomy}
    \vspace{-4mm}
\end{figure*}

\subsection{Task Relation} \label{tr}
Considering the relation of tasks, MTDRS can be systematized as: parallel, cascaded, and auxiliary with main tasks. {Table \ref{tab:relation-summary} gives a compact comparison of these relations to complement the detailed cascaded-model table.}

\begin{table*}[t]
\setlength\abovecaptionskip{0\baselineskip}
\setlength\belowcaptionskip{0\baselineskip}
\centering
\caption{{Summary of task-relation categories in MTDRS}}
\label{tab:relation-summary}
\resizebox{0.98\textwidth}{!}{%
\begin{tabular}{p{0.15\textwidth}p{0.25\textwidth}p{0.28\textwidth}p{0.22\textwidth}}
\toprule
{\textbf{Task Relation}} & {\textbf{Typical Setting}} & {\textbf{Representative Models}} & {\textbf{Main Challenge}} \\ \midrule
{Parallel} & {Multiple tasks are predicted from the same impression or interaction record without an explicit label-order constraint.} & {RnR \cite{hadash2018rank}, MTER \cite{wang2018explainable}, CAML \cite{chen2019co}, DUPN \cite{ni2018perceive}, MRAN \cite{zhao2019multiple}, MSSM \cite{ding2021mssm}, CFS-MTL \cite{chen2022cfs}, ORCA \cite{luo2025orca}} & {Selecting useful shared features while avoiding negative transfer across weakly related tasks.} \\ \midrule
{Cascaded} & {Later behaviors depend on earlier behaviors, such as impression $\rightarrow$ click $\rightarrow$ conversion.} & {ESMM \cite{ma2018entire}, AITM \cite{xi2021modeling}, APEM \cite{tao2023task}, DCMT \cite{zhu2023dcmt}, AECM \cite{zhang2024aecm}, DDPO \cite{su2024ddpo}} & {Modeling sequential dependence under sample selection bias, data sparsity, delayed feedback, and causal confounding.} \\ \midrule
{Auxiliary with Main Task} & {Auxiliary tasks provide extra supervision for a primary task and may be discarded or down-weighted at inference.} & {ESDF \cite{wang2020delayed}, DMTL \cite{zhao2021distillation}, MetaBalance \cite{he2022metabalance}, MTRec \cite{li2020multi}, PICO \cite{lin2022personalized}, Cross-Distill \cite{yang2022cross}} & {Designing auxiliary tasks to help the main task without introducing task dominance or noisy transfer.} \\
\bottomrule
\end{tabular}}
\vspace{-3mm}
\end{table*}

\subsubsection{Parallel} \label{sec:3-2-1}

A parallel task relation indicates that various tasks are independently calculated without the sequential dependency of their results.
The objective function of parallel task relation MTDRS is usually defined as the weighted sum of losses with constant loss weights.
For exposition, existing methods under parallel task relation can be discussed from their target tasks and their modeling challenges.
As for the target, RnR \cite{hadash2018rank} merges the ranking and rating prediction tasks along with a two-phase decision process for personalized recommendation in video recommendation.
Additionally, MTER \cite{wang2018explainable} and CAML \cite{chen2019co} focus on recommendation and explanation tasks. 
Besides, DINOP \cite{xin2019multi} is proposed specifically for e-commerce online promotions by considering multiple sales prediction tasks. 


Meanwhile, several works try to tackle the challenge of feature selection and sharing among parallel tasks, since a static sharing strategy may fail to capture the complex task relevance.
To be specific, existing studies mainly adopt attention mechanisms.
DUPN \cite{ni2018perceive} integrates  multi-task learning, attention along with RNNs to extract general features that will be shared among the associated tasks.
MRAN \cite{zhao2019multiple} proposes to use an attention mechanism for feature interaction and task-feature alignment.
RevMan \cite{li2021revman} uses an attentive adaptive feature sharing mechanism for different tasks.
MSSM \cite{ding2021mssm} applies a feature-field sparse mask within the input layer and the connection control between a set of more fine cells in sub-networks of multiple layers.
Recently, CFS-MTL \cite{chen2022cfs} proposes to select the stable causal features via pseudo-intervention from a causal view. {Recent parallel-task studies further emphasize task decoupling and bottom-level representation specialization. For example, task information decoupling explicitly separates task-related and task-irrelevant information to reduce interference \cite{yan2024multi}, while embedding-level designs such as STEM \cite{su2023stem} and task-specific interaction networks such as DTN \cite{bi2024dtn} indicate that negative transfer can already occur before high-level expert routing. For dwell-time prediction, ORCA \cite{luo2025orca} studies over-reliance among tasks and applies causal decoupling to reduce harmful dependence, making it directly relevant to MTDRS rather than merely general causal transfer learning.}

\begin{table*}[t]
\setlength\abovecaptionskip{0\baselineskip}
\setlength\belowcaptionskip{0\baselineskip}
\centering
\caption{Summary of models with cascaded task relation in MTDRS}
\label{tab:cas}
\resizebox{0.99\textwidth}{!}{%
\begin{tabular}{ccc}
\toprule
\textbf{Model} & \textbf{Problem} & \textbf{Behavior Sequence} \\ \midrule

ESMM \cite{ma2018entire} & SSB \& DS & impression $\rightarrow$ click $\rightarrow$ conversion \\ \midrule

ESM\textsuperscript{2} \cite{wen2020entire} & SSB \& DS & impression $\rightarrow$ click $\rightarrow$ D(O)Action $\rightarrow$ purchase \\ \midrule

Multi-IPW \& DR \cite{zhang2020large} & SSB \& DS & exposure $\rightarrow$ click $\rightarrow$ conversion \\ \midrule

ESDF \cite{wang2020delayed} & SSB \& DS \& time delay & impression $\rightarrow$ click $\rightarrow$ pay \\ \midrule

HM\textsuperscript{3} \cite{wen2021hierarchically} &  SSB \& DS \& micro and macro behavior modeling & impression $\rightarrow$ 
click $\rightarrow$ micro $\rightarrow$ macro $\rightarrow$ purchase \\ \midrule


AITM \cite{xi2021modeling} & sequential dependence in multi-step conversions & impression $\rightarrow$ click $\rightarrow$ application $\rightarrow$ approval $\rightarrow$ activation \\ \midrule

MLPR \cite{wu2022multi} & sequential engagement \& vocabulary mismatch in product ranking & impression $\rightarrow$ click $\rightarrow$ add-to-cart $\rightarrow$ purchase \\ \midrule

ESCM\textsuperscript{2} \cite{wang2022escm2} & inherent estimation bias \& potential independence priority & impression $\rightarrow$ click $\rightarrow$ conversion \\ \midrule

HEROES \cite{jin2022multi} & multi-scale behavior \& unbiased learning-to-rank & observation $\rightarrow$ click $\rightarrow$ conversion \\ \midrule

APEM \cite{tao2023task} & sample-wise representation learning in SDMTL & impression $\rightarrow$ click $\rightarrow$ authorize $\rightarrow$ conversion \\ \midrule

DCMT \cite{zhu2023dcmt} & SSB \& DS \& potential
independence priority (PIP) & exposure $\rightarrow$ click $\rightarrow$ conversion \\ \midrule

{AECM \cite{zhang2024aecm}} & {SSB \& causal debiasing for post-click CVR} & {impression $\rightarrow$ click $\rightarrow$ conversion} \\ \midrule

{DDPO \cite{su2024ddpo}} & {dual propensity debiasing for post-click CVR} & {impression $\rightarrow$ click $\rightarrow$ conversion} \\
\bottomrule

\end{tabular}%
}
\vspace{-5mm}
\end{table*}

\subsubsection{Cascaded} \label{casc}
A cascaded task relationship refers to the sequential dependency between tasks. In other words, the computation of the current task depends on the previous ones, e.g., CTCVR derived by multiplying the predicted CTR and CVR probabilities in the ESMM-style decomposition. It can be considered as a general MTL problem with the assumption on prediction scores for specific task $k$:
\begin{equation}
    \hat{y}_n^k(\theta^s, \theta^k) - \hat{y}_n^{k-1}(\theta^s, \theta^k) = P(\epsilon_k=0, \epsilon_{k-1}=1)
\end{equation}
where $\epsilon_k$ is the indicator variable for task $k$. This assumption implies the difference of $\hat{y}_n^k(\theta^s,\theta^k)$ and $\hat{y}_n^{k-1}$ is the probability of the task $k$ not happening while the task $k-1$ is observed. Besides, the formulation above is equivalent to the sequential dependence MTL (SDMTL) in \cite{tao2023task}. {In most cascaded recommendation settings, the labels are binary behavior indicators, but extensions may replace them with calibrated probabilities, delayed conversion labels, or continuous post-click utilities.}

We summarize the methods under the cascaded task relation setting in Table \ref{tab:cas} with respect to the problem and behavior sequence. They mainly aim at CVR prediction task on the e-commerce platform, except AITM \cite{xi2021modeling} and APEM \cite{tao2023task} are proposed for advertising and financial service, respectively. Meanwhile, they mainly target at tackling sample selection bias (SSB) and data sparsity (DS) issues caused by sparse training data of conversion. Besides, their assumed sequential patterns are based on ``impression $\rightarrow$ click $\rightarrow$ conversion'' and its extension following ESMM \cite{ma2018entire}, which adopts shared embedding and models over entire space. {After 2023, causal debiasing remains active in this line. AECM \cite{zhang2024aecm} introduces an adversarial-enhanced causal multi-task framework for post-click CVR debiasing, while DDPO \cite{su2024ddpo} directly optimizes dual propensities for post-click conversion estimation. These works are included here because their core problem remains cascaded multi-task conversion estimation rather than pure multi-scenario recommendation.}

\subsubsection{Auxiliary with Main Task} \label{atl}
Auxiliary with main task refers to the circumstance that a task is specified as the main task while others, i.e., associated auxiliary tasks help to improve its performance.
The probability estimation for the main task is calculated based on the probability of auxiliary tasks, which is estimated on the entire space with richer information.
On the one hand, some works simply adopt the original recommendation tasks as auxiliaries \cite{zhang2020large,wang2020delayed,zhao2021distillation,he2022metabalance}. Specifically, Multi-IPW and Multi-DR \cite{zhang2020large} introduce an auxiliary CTR task with main CVR and imputation task. ESDF \cite{wang2020delayed} treats CTR and CTCVR as auxiliaries of time delay task. DMTL \cite{zhao2021distillation} models CTR as auxiliary of duration task. 

On the other hand, some works design various auxiliary tasks under specific settings \cite{li2020multi,lin2022personalized,yang2021multi,yang2022cross}. Specifically, MTRec \cite{li2020multi} takes link prediction for network dynamic modeling as an auxiliary of the recommendation task. PICO \cite{lin2022personalized} considers task relevance between CTR and CVR as auxiliary. MTAE \cite{yang2021multi} predicts the winning probability as auxiliary. Cross-Distill \cite{yang2022cross} proposes ranking-based task as auxiliary containing cross-task relation. Specially, CSRec \cite {bai2022contrastive} and PICO \cite{lin2022personalized} adopt contrastive learning as the auxiliary task to extract task relevance better.
Nevertheless, the frameworks above are manually-auxiliary, since the design of auxiliary tasks usually requires specific domain knowledge. Recently, Wang et al. \cite{wang2022can} propose under-parameterized self-auxiliaries to achieve better generalization.

\subsection{Methodology} \label{sec:3-3}
For methodology, existing MTDRS can be categorized into: parameter sharing, optimization, and training mechanism.

\begin{figure*}[t]
    \setlength\abovecaptionskip{0.3\baselineskip}
    \setlength\belowcaptionskip{0.1\baselineskip}
	\centering
	\includegraphics[scale=0.3,trim=0 7 0 0,clip]{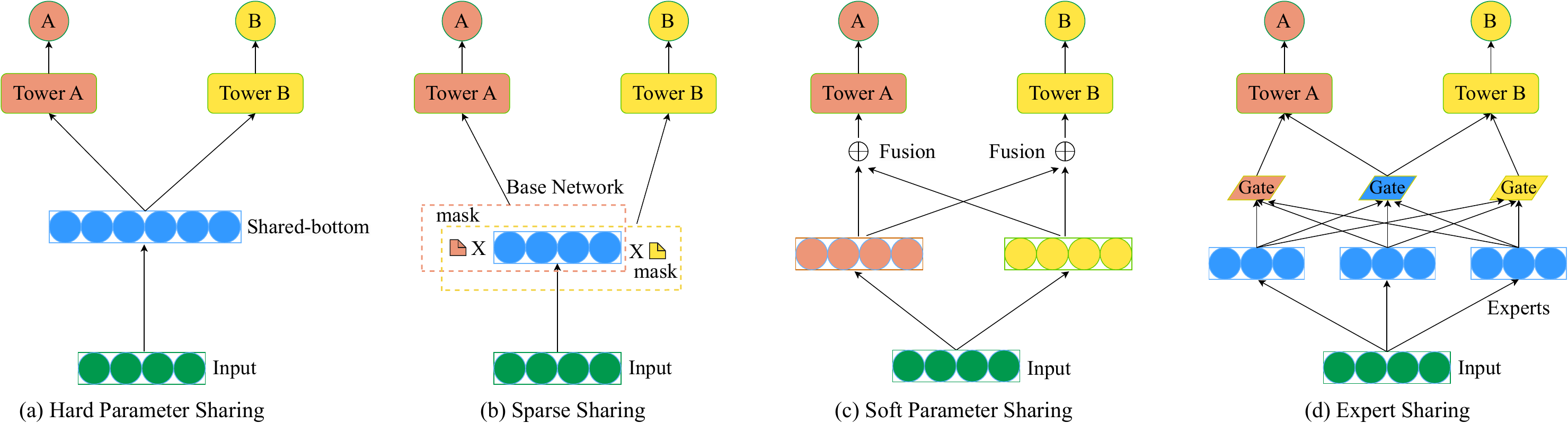}
	\caption{The illustration of hard sharing, sparse sharing, soft sharing, and expert sharing with task A and B. Blue represents the shared parameters. Red and yellow represent the task-specific parameters.}
	\label{parameter-sharing}
\vspace{-5mm}
\end{figure*}

\subsubsection{Parameter Sharing} \label{sec:3-3-1}
Considering the parameter sharing pattern and sharing levels, existing methods can be categorized into hard sharing, sparse sharing, soft sharing, and expert sharing. Their illustrations are shown in Figure \ref{parameter-sharing} with two tasks as an example.

\begin{itemize}[leftmargin=*]

\item \textbf{Hard Sharing.}
Hard parameter sharing refers to the shared bottom layers extracting the same information for different tasks, while the task-specific top layers are trained individually. On the one hand, it can improve computation efficiency and alleviate over-fitting. However, it may suffer from the limited capacity of the shared parameter space, especially for weakly related tasks and noise.
To be specific, many studies simply follow the shared bottom structure of hard parameter sharing \cite{liu2022multi,he2022metabalance,yang2022adatask}. Notably, MFH \cite{liu2022multi} proposes an efficient cooperative learning with heterogeneity, which consists of nested hierarchical MTL trees for multi-dimension task relation from the macro perspective and switcher networks in different levels from the micro perspective. 


\item \textbf{Sparse Sharing.}
Sparse sharing extracts sub-networks for each task by parameter masks from a base network in the shared parameter space.
It is a special case of hard sharing because the parameters are shared in neuron or layer level.
Specifically, it usually extracts subnets and trains with the fixed sharing strategy in parallel for each task.
Compared with hard sharing, it has the advantage of coping with the weakly related tasks flexibly.
 In MTDRS, LT4REC \cite{xiao2020lt4rec} proposes neuron-level mask based on Lottery Ticket Hypothesis.
MSSM \cite{ding2021mssm} adopts sparse connection in field-level for task-aware feature selection and in cell-level for determining the connection among cells in subnets of different layers.
However, these methods may suffer from negative transfer when updating shared parameters.
Recently, CSRec \cite{bai2022contrastive} proposes contrastive pruning in contrastive sharing network to learn better parameter masks among tasks. {The broader sparse-MoE literature also warns that routing can induce representation collapse by clustering representations around expert centroids \cite{chi2022representationcollapse}. In recommendation, embedding collapse has also been observed when scaling recommendation models \cite{guo2024embeddingcollapse}. Although these observations are not specific to multi-task recommendation, they are relevant to MTDRS considering model architectures and scaling of MTDRS models.}


\item \textbf{Soft Sharing.}
In contrast to the hard sharing method extracting information from the shared bottom layer, soft parameter sharing also builds separate models for tasks but the information among tasks is fused by weights of task relevance, e.g., from the attention mechanism. This structure could equip the model with relatively high flexibility in parameter sharing compared to hard sharing methods. However, {it cannot fully reconcile model flexibility with computation cost.}
In MTDRS, DINOP \cite{xin2019multi} incorporates dynamic target user profiles into the attention-based pooling, so as to learn universal item representation for different tasks. 

\item \textbf{Expert Sharing.}
Expert sharing model first employs multiple expert networks to extract knowledge from the shared bottom layer. Then the knowledge is fed into task-specific modules like gates to learn useful information. Finally, the assembled information is passed into the task-specific tower. Notably, expert sharing is a special case of soft sharing since expert knowledge is shared but fused by task-specific weights.  

\vspace{0.4em}

Inspired by Mixture of Experts (MoE) \cite{jacobs1991adaptive}, MMoE \cite{ma2018modeling} proposes to use softmax gates from gating networks to assemble experts with different weights for each task, and it acts as a milestone in MTDRS. Based on MMoE, further improvements have been proposed subsequently. 
SNR \cite{ma2019snr} {proposes} to split the shared bottom layer into sub-networks allowing sparse connection and replace the gating network with latent variable.
PLE \cite{tang2020progressive} utilizes a novel Customized Gate Control (CGC) module which could explicitly separate shared and task-specific experts. Meanwhile, it adopts multi-level extraction networks with progressive separation routing.

\vspace{0.4em}

Afterward, more works have been proposed based on MMoE. 
DMTL \cite{zhao2021distillation} proposes to distill information to a student model learned from the teacher model, which employs the MMoE framework to model CTR and CVR. DSelect-k \cite{hazimeh2021dselect} proposes a continuously differentiable sparse gate to tackle the lack of smoothness. Meanwhile, MetaHeac \cite{zhu2021learning} introduces a hybrid structure that combines a critic network with an expert network. Recently, PFE \cite{xin2022prototype} adapts prototype feature extraction into the MMoE framework. MVKE \cite{xu2022mixture} proposes a virtual kernel structure for expert and gate for better user profiling. FDN \cite{zhou2023feature} decomposes features with constraints to tackle feature redundancy. AdaTT \cite{li2023adatt} uses gating and residual mechanism for task-adaptive fusion of expert. STEM \cite{su2023stem} adopts shared and task-specific embeddings equipped with customized gating network. DTN \cite{bi2024dtn} observes the divergent importance values of the same feature across tasks, and {proposes} to use various feature interactions modules and task-sensitive network. MoME \cite{xu2024mome} proposes to adopt an over-parameterized base neural network to train an MoE by model pruning and expert masking. {These recent models show a shift from simply increasing the number of experts to controlling where sharing happens: embeddings, feature interactions, experts, gates, and task towers may all need task-aware design.}
MoSE \cite{qin2020multitask} models sequential user behaviors by using sequential experts and applying Long Short-Term Memory (LSTM) to the shared bottom layer and task-specific towers. 

\begin{table*}[t]
\setlength\abovecaptionskip{0\baselineskip}
\setlength\belowcaptionskip{0\baselineskip}
\centering
\caption{{Critical comparison of representative methodology families in MTDRS}}
\label{tab:methodology-comparison}
\resizebox{0.98\textwidth}{!}{%
\begin{tabular}{p{0.17\textwidth}p{0.24\textwidth}p{0.24\textwidth}p{0.24\textwidth}}
\toprule
{\textbf{Family}} & {\textbf{Strength}} & {\textbf{Limitation}} & {\textbf{Industrial Consideration}} \\ \midrule
{Hard sharing} & {Efficient serving and simple deployment with one shared representation.} & {Limited capacity for weakly related tasks and prone to seesaw effects.} & {Attractive for latency-sensitive ranking, but requires careful task selection.} \\ \midrule
{Sparse sharing} & {Allocates partial sub-networks or masks to tasks, improving flexibility.} & {Mask search and routing can be unstable, and sparse experts may suffer representation collapse.} & {Useful when model size is constrained, but online routing and pruning need monitoring.} \\ \midrule
{Soft/expert sharing} & {Learns task-adaptive mixtures and captures non-uniform task relatedness.} & {More experts increase memory, gate computation, and risk of over-specialization.} & {Widely used in large-scale systems, but gate latency and expert utilization must be controlled.} \\ \midrule
{Optimization-based methods} & {Directly address gradient domination, conflict, and loss imbalance.} & {Often improve training dynamics without changing serving architecture, but may require sensitive hyper-parameters.} & {Convenient when online architecture cannot be changed, yet offline-online metric mismatch remains.} \\ \midrule
{RL-based fusion} & {Optimizes long-term satisfaction after task predictions are produced.} & {Requires reliable reward design, off-policy correction, and exploration control.} & {Promising for final ranking fusion, but safety, delayed reward, and real-time constraints are critical.} \\
\bottomrule
\end{tabular}}
\vspace{-5mm}
\end{table*}

\end{itemize}

\subsubsection{Optimization} \label{sec:3-3-2}

From the optimization aspect, in MTL, the joint optimization of different tasks 
needs to tackle two issues: (i) negative transfer among tasks, where improving one task can degrade another due to harmful shared gradients or representations, and (ii) multi-objective trade-off, where a deployed recommender must balance objectives such as accuracy, revenue, fairness, diversity, or long-term satisfaction. {Thus, ``conflict'' here refers to training interference among task losses, while ``trade-off'' refers to the preference relation among business or recommendation objectives after task predictions are obtained.}
\begin{itemize}[leftmargin=*]
\item \textbf{Negative Transfer.}\label{nt}
Negative transfer is a situation that transferring unrelated information among tasks could result in performance degradation and seesaw phenomenon. When the performance of some tasks is improved but at the cost of others' result, the seesaw phenomenon is observed. Since most of the existing methods \cite{yan2024multi} seek better recommendation accuracy, they focus on this problem. Specifically, there are mainly two reasons causing negative transfer in MTDRS from the shared parameters $\theta$. The first is gradient dominating and the second is parameter conflict.


Gradient dominating denotes the magnitude imbalance of gradient $\Vert \nabla_{\theta} L^k(\theta) \Vert$ of different tasks, and some works try to tackle this problem \cite{chen2018gradnorm,yu2020gradient}. In the recommendation community, AdaTask \cite{yang2022adatask} proposes to {quantify} task dominance of shared parameters, and calculating task-specific accumulative gradients for adaptive learning rate methods. 
MetaBalance \cite{he2022metabalance} proposes to flexibly balance the gradient magnitude proximity between auxiliary and target tasks by a relax factor.


Besides, parameter conflict indicates that the shared parameter $\theta$ has opposite directions of gradient $\nabla_{\theta} L^k(\theta)$ in different tasks.
PLE \cite{tang2020progressive} discusses seesaw phenomenon and propose Customized Gate Control (CGC) that separates shared and task-specific experts to explicitly alleviate parameter conflicts.
CSRec \cite{bai2022contrastive} applies an alternating training procedure and contrastive learning on parameter masks to reduce the conflict probability.
PEPNet \cite{chang2023pepnet} proposes dynamic network to tackle the imperfectly double seesaw phenomenon between domains and tasks.
Meanwhile, GradCraft \cite{bai2024gradcraft} simultaneously tackles these two problems. They propose to adjust gradient norm and deconflict global direction through projection and combination.
{Beyond gradient-level remedies, recent task-decoupling methods argue that negative transfer can be rooted in representation entanglement. TIDM \cite{yan2024multi} decouples task information to reduce interference, while STEM \cite{su2023stem}, DTN \cite{bi2024dtn}, and MoME \cite{xu2024mome} address task-specificity at embedding, feature-interaction, and expert-mask levels. These developments suggest that representation collapse and task collapse should be treated as first-class risks in expert-based MTDRS.}


\item \textbf{Multi-objective Trade-off.}\label{mt}
The trade-offs among objectives under {the} MTDRS setting {are} a new topic. Specifically, the corresponding objectives in each task are usually optimized by a single model regardless of the potential conflict.
Some study the trade-off between group fairness and accuracy across multiple tasks \cite{wang2021understanding} and afterward, the trade-off between minimizing task conflicts and improving multi-task generalization in a higher level \cite{wang2022can}.

\end{itemize}

\subsubsection{Training Mechanism} \label{md}
Training mechanism refers to the specific training process and learning strategy of different tasks in MTDRS model. Existing works on MTDRS can be grouped into: joint training, reinforcement learning, and auxiliary task learning. {Joint training means that task losses are optimized within one training process over shared and task-specific parameters, although the batches, loss weights, or task schedules may vary across methods.}

\begin{itemize}[leftmargin=*]
\item \textbf{Joint Training.}
Most MTL models adopt joint training among tasks in a parallel manner, and the majority of the above-mentioned MTDRS models belong to this category. Specially, some works simply jointly learn different tasks, 
such as session-based RS \cite{shalaby2022m2trec,qiu2021incorporating,meng2020incorporating}, route RS \cite{das2022marrs}, knowledge graph enhanced RS \cite{wang2019multi}, explainability \cite{lu2018like,wang2018explainable}, and graph-based RS \cite{wang2020m2grl}. Besides, some works adopt an alternating training procedure, e.g., contrastive pruning \cite{bai2022contrastive}.

\item \textbf{Reinforcement Learning.}\label{rl}
Reinforcement Learning (RL) algorithms have recently been applied in DRS, which models the sequential user behaviors as Markov Decision Process (MDP) and utilizes RL to generate recommendations at each decision step \cite{mahmood2007learning}. By setting user-item features as state and continuous score pairs for multiple tasks as actions, the RL-based MTL method is capable of handling the sequential user-item interaction and optimizing long-term user engagement. Zhang et al. \cite{zhang2022multi} formulate MTF as MDP and use batch RL to optimize long-term user satisfaction. Han et al. \cite{han2019optimizing} use an actor-critic model to learn the optimal fusion weight of CTR and the bid rather than greedy ranking strategies to maximize the long-term revenue. Liu et al. \cite{liu2023multi} use dynamic critic networks to adaptively adjust the fusion weight considering the session-wise property. Recently, to tackle the overstrict constraints and inefficiency, Tencent proposes to leverage off-policy RL with exploration policy \cite{liu2024off} which improves RL-MTF. {EnhancedRL \cite{liu2024enhancedrl} expands the RL state from user-level features to user--item and contextual features, while xMTF \cite{cao2025xmtf} replaces hand-crafted fusion formulas with learnable monotonic fusion cells. These post-2023 works indicate that RL-based MTF is moving from coefficient tuning over fixed formulas toward richer state modeling and more flexible fusion-function learning.}

\item \textbf{Auxiliary Task Learning.}
As discussed in Section \ref{atl}, adding auxiliary tasks {aims to enhance} the performance of primary tasks. Specifically, the auxiliary tasks are usually trained along with the primary tasks in a joint training manner. By contrast, ESDF \cite{wang2020delayed} employs Expectation-Maximization
(EM) algorithm for optimization. Besides, self-auxiliaries are trained with task-specific sub-networks while they are discarded in the inference stage.
\end{itemize}

%% file: 3Application.tex
\section{Application and Datasets} \label{app}
In this section, the application of MTDRS on different fields and multi-task fusion is first introduced. Next, the public datasets of MTR are summarized. {We focus on applications where multiple recommendation tasks or feedback labels are learned jointly, and we use datasets only as representative public resources rather than an exhaustive benchmark list.}

\subsection{Application Fields}
Apart from e-commerce, MTDRS has been applied to various fields such as advertising and social media. On the one hand, for an advertisement, to jointly and accurately estimate the utility and cost are essential for determining its success. In MTDRS, MM-DFM \cite{hou2021conversion} performs multiple conversion prediction tasks in different observation duration. MetaHeac \cite{zhu2021learning} is proposed for audience expansion tasks on content based mobile marketing. MVKE \cite{xu2022mixture} is proposed for user tagging for online advertising. {Recent causal conversion models such as AECM \cite{zhang2024aecm} and DDPO \cite{su2024ddpo} further show that industrial advertising systems need to jointly model click and post-click labels while correcting exposure, click, and conversion biases.}
On the other hand, social media is a more complex field since the users interact with both items and users. Multiple MTDRS models validate their effectiveness on social media by online A/B test, including MMoE \cite{zhao2019recommending} on YouTube considering engagement and satisfaction, LT4REC \cite{xiao2020lt4rec} on Tencent Video, and BatchRL-MTF \cite{zhang2022multi} on Tencent short video platform. {For short-video and content feeds, MTDRS is often connected to long-term engagement objectives, where click, watch time, like, share, comment, and follow signals must be fused under serving-latency and online-exploration constraints.}

\subsection{Multi-task Fusion}
In real-world multi-task RS, after predicting various recommendation tasks, a multi-task fusion (MTF) model is usually applied to combine the multi-task outputs into one ranking score considering user satisfaction and produce the final ranking. 
Several works try to search for the fusion weight, such as Grid
Search \cite{gu2020deep,rodriguez2012multiple}, Evolutionary Algorithm \cite{ribeiro2014multiobjective}, and Bayesian Optimization \cite{galuzzi2020hyperparameter}. A state-of-the-art solution is RL \cite{han2019optimizing,zhang2022multi,liu2023multi,liu2024off,liu2024enhancedrl,cao2025xmtf}, which is discussed in Section \ref{rl}. {Compared with static linear fusion, recent RL-based MTF methods explicitly consider delayed user satisfaction and dynamic context, but they also introduce production risks such as off-policy evaluation error, exploration safety, reward hacking, and additional real-time inference cost.}

\begin{table*}[t]
\setlength\belowcaptionskip{0\baselineskip}
\setlength\abovecaptionskip{0\baselineskip}
\centering
\caption{Summary of datasets for MTDRS}
\label{tab:sum}
\resizebox{1\textwidth}{!}{%
\begin{tabular}{cccc}
\toprule
\textbf{Datasets} & \textbf{Stage} & \textbf{Tasks} & \textbf{Website} \\ \midrule

Ali-CCP \cite{ma2018entire} & Ranking & CTR, CVR & \url{https://tianchi.aliyun.com/dataset/408/}  \\ \midrule
Criteo \cite{diemert2017attribution} & Ranking & CTR, CVR & \url{https://ailab.criteo.com/criteo-attribution-modeling-bidding-dataset/} \\ \midrule
AliExpress \cite{li2020improving} & Ranking & CTR, CTCVR & \url{https://tianchi.aliyun.com/dataset/74690/}  \\ \midrule
MovieLens \cite{harper2015movielens} & Recall \& Ranking & Watch, Rating & \url{https://grouplens.org/datasets/movielens/}  \\ \midrule
Yelp & Recall \& Ranking & Rating, Explanation & \url{https://www.yelp.com/dataset/} \\ \midrule
Amazon \cite{he2016ups} & Recall \& Ranking & Rating, Explanation & \url{http://jmcauley.ucsd.edu/data/amazon/} \\ \midrule
Kuairand \cite{gao2022kuairand} & Recall \& Ranking & Click, Like, Follow, Comment, \dots & \url{https://kuairand.com/} \\ \midrule
Tenrec \cite{yuan2022tenrec} & Recall \& Ranking & Click, Like, Share, Follow, \dots & \url{https://github.com/yuangh-x/2022-NIPS-Tenrec/} \\ \bottomrule
\end{tabular}}

\vspace{-3mm}
\end{table*}

\subsection{Datasets} \label{data}
We summarize the public datasets in MTDRS in Table \ref{tab:sum} according to their stage of recommendation applied and the used frequency. {The dataset list prioritizes public resources that have been repeatedly used for multi-task, multi-behavior, conversion, explanation, or generative recommendation experiments. Our inclusion criteria are: (i) the dataset is publicly accessible; (ii) it contains native multiple recommendation labels or has been widely used in MTDRS papers with a clearly specified multi-task construction; and (iii) the tasks can be interpreted as recommendation tasks rather than only side information. We do not aim to enumerate all datasets that can be post-processed into a multi-task benchmark. For example, music-streaming logs or other consumption datasets may contain play counts, repeated listening, ratings, or skips, and one can derive implicit feedback by thresholding or counting repeated consumption. However, if the released dataset does not provide a standard multi-task split or has not been commonly adopted in MTDRS studies, including it would make the benchmark list difficult to reproduce and would mix native multi-task datasets with researcher-defined task transformations. Therefore, we report representative datasets with explicit or commonly used MTDRS task definitions, while acknowledging that additional domain datasets can be adapted when researchers clearly document the derived labels and preprocessing protocol.} Meanwhile, it is notable that the ratings can be categorized into binary classification task, i.e., to predict whether a rating ranging from 1 to 5 is greater than 3. For MovieLens dataset, the common two tasks are to predict whether the user watches the movie and to predict the rating. 

%% file: 5FutureDirections.tex
\section{Challenges and Future Directions} \label{fd}
We summarize the challenges and future directions of MTDRS including negative transfer, multi-task with multi-scenario modeling, using large pre-trained model, AutoML, explainability, {task-specific biases, and industrial deployment}.

\begin{itemize}[leftmargin=*]
\item \textbf{Negative Transfer.}
As discussed in Section \ref{nt}, previous works try to tackle negative transfer in MTDRS from either gradient or separating shared and specific parameters. However, how to extract the complex inter-task correlation needs further research, e.g., from the causal relation \cite{chen2022cfs}. Meanwhile, what, where, and when to transfer to alleviate negative transfer is still under-explored in MTDRS. {Recent studies suggest that negative transfer is not only a loss-balancing problem, but also a representation problem. Task-specific embeddings, feature-interaction modules, expert masks, and information decoupling \cite{su2023stem,bi2024dtn,xu2024mome,yan2024multi} can reduce task interference before gradients reach the shared tower. ORCA \cite{luo2025orca} further indicates that causal decoupling can mitigate over-reliance in multi-task dwell-time prediction. A related risk is representation or embedding collapse, where representations lose useful task- or item-level diversity; this has been studied in sparse expert systems \cite{chi2022representationcollapse} and large recommendation models \cite{guo2024embeddingcollapse}. Future MTDRS should therefore diagnose collapse, task dominance, and expert under-utilization together instead of reporting only per-task accuracy.}

\item \textbf{Multi-task with Multi-scenario Modeling.}
There is a new trend to tackle multi-task and multi-scenario problems in a unified model in an end-to-end manner \cite{chang2023pepnet,zhou2023hinet,lan2023m3rec,zhang20233mn,yi2025adaptive}. For example, 
M2M \cite{zhang2022leaving} proposes to use meta learning to extract specific information from scenario knowledge as dynamic weights considering inter-scenario correlations. AESM\textsuperscript{2} \cite{zou2022automatic} proposes to adaptively select shared and specific experts by calculating relevance score via gating mechanism, and stack multiple layers to model hierarchical scenario structure. However, 
the tasks with sequential dependency are not considered. {A key open problem is how to jointly model scenario shift and task heterogeneity without confusing scenarios with tasks. Pure MSR methods should not be directly counted as MTDRS, but multi-scenario multi-task systems must decide whether sharing happens across tasks, scenarios, or their Cartesian product. This raises new questions about sparse data in long-tail scenarios, scenario-specific label bias, and online serving cost when each request needs both scenario routing and task routing.}

\item \textbf{Using Large Language Model.}
Large language model 
has achieved great success, and it is promising to conduct MTR by the large pre-trained language model, which is able to unify different recommendation tasks in a single sequence-to-sequence framework. {This line has evolved from pre-trained language model based recommenders to LLM-based universal recommendation learners.} Geng et al. \cite{geng2022recommendation} propose P5 with pre-training and personalized hard prompts. It covers five recommendation task families based on T5 backbone \cite{raffel2020exploring}. Similarly, M6-Rec \cite{cui2022m6} is proposed upon M6 \cite{lin2021m6} and UniMIND \cite{deng2022unified} is proposed for multi-goal conversational recommendation. {More recently, URM \cite{jiang2025urm} argues that LLMs can serve as universal recommendation learners by transferring a broad set of recommendation tasks into a unified language modeling interface, showing a clear progression from P5 and M6-Rec-style pre-trained sequence-to-sequence recommendation toward LLM-centered recommendation unification.} However, they all rely on prompt design or tuning for each specific task, which is inefficient. 
{Recent work further extends this direction from prompt-based unification to LLM-assisted multi-task recommendation. CKF \cite{zhao2025ckf} uses LLM-based collaborative knowledge fusion for multi-task recommender systems. Other unified generative recommender models, such as IDGenRec \cite{tan2024idgenrec}, COBRA \cite{yang2025cobra}, and UniGRF \cite{zhang2025unigrf}, are adjacent to MTDRS because they unify recommendation interfaces such as ID generation, retrieval, and ranking rather than explicitly optimizing multiple behavior-label tasks. These models are nevertheless useful for MTDRS because they show how heterogeneous recommendation outputs can be cast into a sequence-generation interface. However, they also bring open challenges: how to align generative objectives with ranking metrics, how to support low-latency serving, how to avoid hallucinated explanations or items, and how to combine dense ID-based collaborative signals with textual world knowledge.}

\item \textbf{AutoML.}
For MTDRS, different tasks may require different neural network architectures and hyper-parameters, and {manually designing and tuning different components for each task is computationally costly}. Recently, automated machine learning (AutoML) \cite{chen2022automated} is emerging for automating the
components of DRS and enhancing generalization. Under MTDRS setting, SNR \cite{ma2019snr} proposes to automatically learn the connection routing for flexible parameter sharing by Neural Architecture Search (NAS). Besides, MTNAS \cite{chen2021boosting} proposes to search for the suitable sparse sharing route for MTDRS by NAS. Jiang et al. \cite{jiang2024automatic} proposes a Progressively Discretizing Differentiable Architecture Search algorithm for efficient exploration in search space. However, they only focus on the parameter sharing routing, while other components and hyper-parameters are still under-explored.


\item \textbf{Explainability.}
Explainable recommendation is proposed to improve the transparency and user satisfaction in trustworthy RS \cite{fan2022comprehensive}. 
Most existing DRS models employ DNN, whose prediction mechanism is difficult to explain, suffering from vulnerability and unreliability. 
MTDRS faces more challenges with explainability due to its complex task relevance. 
Works such as \cite{wang2018explainable,chen2019co} address the issue but only for single score prediction tasks, not multiple recommendation tasks. 
Consequently, it is worth paying more attention to this topic, leading to a better trustworthy MTDRS.

\item \textbf{Task-specific Biases.}
MTDRS may be prone to the task-specific biases, which are caused by different behavior patterns in different tasks. 
They can lead to an unfair or sub-optimal experience for users and result in poor performance on certain tasks.
Most existing models only focus on one specific bias, such as sample selection bias \cite{ma2018entire}, implicit selection bias \cite{zhao2019recommending}, and inherent estimation bias \cite{wang2022escm2}. {Recent causal post-click conversion models \cite{zhang2024aecm,su2024ddpo} indicate that exposure, click, conversion, and observation processes may each introduce different propensities. Consequently, how to tackle biases among different tasks needs further study.}

\item {\textbf{Industrial Deployment.}}
{Industrial MTDRS faces constraints that are less visible in offline benchmarks. First, online serving requires strict latency and memory control. Therefore, complex expert routing, LLM generation, or RL fusion must be compatible with real-time ranking. Second, task labels are delayed and biased; conversion, retention, and satisfaction labels may arrive much later than clicks, making online evaluation and credit assignment difficult. Third, fusion strategies must handle safety and business constraints, because aggressive optimization of one task can reduce user trust, creator ecosystem health, or long-term retention. Finally, production MTDRS needs monitoring for task dominance, expert load imbalance, calibration drift, and offline-online metric gaps.}


\end{itemize}

%% file: 6Conslusion.tex
\section{Conclusion} \label{concl}
Multi-task recommendation (MTR) is an important topic in the recommendation community.
In this survey, we are the first to conduct a comprehensive survey on the existing multi-task approaches in DRS, namely multi-task deep recommender systems (MTDRS). Based on our formulation and comparison with similar directions in recommendations, we categorize MTDRS methods into task relation and methodology and depict its trend. Next, the application and public datasets of MTDRS are summarized. Finally, the challenges and future directions are introduced. We hope this survey would shed light on the future study in MTDRS.

\section*{ACKNOWLEDGEMENT}
This research was partially supported by Huawei (Huawei Innovation Research Program, Huawei Fellowship), Research Impact Fund (No.R1015-23), and Collaborative Research Fund (No.C1043-24GF).
